\documentstyle[12pt]{article}
\newcommand{\gurses}{G\"{u}rses}
\newcommand{\bold}[1]{\mbox{{\bf {#1}}}}
\renewcommand{\cite}[1]{\mbox{$^{#1}$}}

\begin{document}

\title{\gurses' Type (b) Transformations are Neighborhood-Isometries}
\author{Isidore Hauser\thanks{Home address:
	4500 19th Street, No.\ 342, Boulder, CO 80304.}
	and Frederick J.\ Ernst \\
	FJE Enterprises\thanks{A private company supporting
	research of F.\ J.\ Ernst and I.\ Hauser
	in classical general relativity theory.} \\
	Rt.\ 1, Box 246A, Potsdam, NY 13676}
\maketitle
\begin{abstract}
Following an idea close to one given by C.\ G.\ Torre (private
communication), we prove that Riemannian spaces $(M,g)$ and $(M,h)$
that are related by a \gurses\ type (b) transformation
[M.\ \gurses, Phys.\ Rev.\ Lett.\ {\bf 70}, 367 (1993)] or,
equivalently, by a Torre-Anderson generalized diffeomorphism
[C.\ G.\ Torre and I.\ M.\ Anderson, Phys.\ Rev.\ Lett.
{\bf xx}, xxx (1993)] are {\em neighborhood-isometric}, i.e.,
every point $\bold{x}$ in $M$ has a corresponding diffeomorphism
$\phi$ of a neighborhood $V$ of $\bold{x}$ onto a generally
different neighborhood $W$ of $\bold{x}$ such that
$\phi^{*}(h|_{W}) = g|_{V}$.
\end{abstract}

\newpage
\section{Introduction}
There have been two divergent opinions concerning those generalized
symmetries of the Einstein vacuum field equations which were found by M.\
\gurses\cite{1} and which he designated as {\em type (b)}.  One opinion,
which both \gurses\ and the authors once entertained, was that some type
(b) transformations can be used (at least in principle) to generate new
exact solutions.  An opinion which suggests the opposite was given by
C.\ G.\ Torre and I.\ M.\ Anderson\cite{2} in their analysis of
generalized symmetries.  They stated that the {\em generalized
diffeomorphism symmetries} which they discovered are physically trivial,
and we shall see\cite{3} that a type (b) generator is equal to a
generalized diffeomorphism generator apart from an inconsequential Lorentz
transformation term.  The reason for their opinion has been explained to
us by Torre\cite{4} and it is our efforts to place his explanation on
secure mathematical ground that has led to this paper.

The objectives of this paper are to introduce the concept of {\em
neighborhood-isometric\cite{5} spacetimes} (an explication of the
description furnished to us by Torre) and to prove that spacetimes which
are related by a type (b) transformation are neighborhood-isometric.
Before we do that, we shall define type (b) transformations in a way which
will facilitate our proof.  The vacuum condition will be ignored in our
definition because, as we shall further stress in the discussions of
Sec.\ VII, this condition is not required for the proof of our theorem.
In Sec.\ VII, we shall also explain our own viewpoint on whether
spacetimes which are related by a type (b) transformation are physically
indistinguishable.  The theorem which we shall prove implies that no type
(b) transformation can be used to generate a new exact analytic solution.
On the other hand, we shall state a theorem which remains to be proved
before one can assert correctly and without reservation that spacetimes
which are related by a type (b) transformation are physically equivalent.

\section{Definition of the Type (b) Transformations}

Our definition of the type (b) 1-parameter family of transformations
differs from but is equivalent to that of \gurses.\cite{1}  The
equivalence and the correspondences with his notations will be detailed in
a lengthier paper\cite{6} which will cover all of his transformation
types.

The parameter will be denoted by $\epsilon$.  (\gurses\ uses
`$\epsilon_{0}$'.)  At $\epsilon=0$, suppose one prescribes the following
objects:
\begin{enumerate}
\item
A Spacetime $(M,g(0))$.
\item
An orthonormal tetrad of 1-forms $e^{b}(0) \; (b=1,2,3,4)$ with the domain
$M$.  It is granted that $M$ is restricted so that a tetrad exists.  The
manifold and all prescribed functions are assumed to be
$\bold{C}^{\infty}$.
\item
A vector field $\bold{a}(0)$ whose domain is $M$.
\end{enumerate}
Then the type (b) transformation
$$\left(M,g(0),\{e^{b}(0)\},\bold{a}(0)\right) \rightarrow
\left(M,g(\epsilon),\{e^{b}(\epsilon)\},\bold{a}(\epsilon)\right)$$
is defined as follows:
\begin{description}
\item[Definition:]  First define the real-valued fields $a^{b}$ by the
equation
\begin{equation}
\bold{a}(0) = a^{b} \bold{e}_{b}(0)
\end{equation}
where $\{\bold{e}_{b}(0)\}$ is the dual basis of $\{e^{b}(0)\}$.
(All p-vectors are denoted by boldface letters, and all p-forms by
lightface.)  Let $e^{b}(\epsilon)$ denote the integrals over a maximally
extended connected interval $J \subset R^{1}$ of the family of equations
\begin{equation}
\partial e^{b}(\epsilon)/\partial \epsilon = - da^{b}
- a^{c} \Gamma_{cd}(\epsilon) \eta^{db}
\end{equation}
where the connection 1-forms $\Gamma_{cd}(\epsilon)$ are defined in terms
of $e^{b}(\epsilon)$ by the familiar structural equations
\begin{equation}
d \wedge e^{b}(\epsilon) = e^{c}(\epsilon) \wedge \Gamma_{cd}(\epsilon)
\eta^{db} \; , \; \Gamma_{cd}(\epsilon) = - \Gamma_{dc}(\epsilon) \; ,
\end{equation}
and $\eta^{db} := 0$ if $d \ne b$ and $\eta^{11} = \eta^{22} = \eta^{33}
= - \eta^{44} := 1$.  Also, let
\begin{equation}
g(\epsilon) := \eta_{ab} e^{a}(\epsilon) \otimes e^{b}(\epsilon)
\end{equation}
where $\eta^{ab}\eta_{bc}=\delta^{a}_{c}$.  Thereupon, $(M,g(\epsilon))$
is automatically a spacetime for which $\{e^{b}(\epsilon)\}$ is an
orthonormal tetrad.  Upon letting $\{\bold{e}_{b}(\epsilon)\}$ be the dual
basis of $\{e^{b}(\epsilon)\}$ and
\begin{equation}
\bold{a}(\epsilon) := a^{b} \bold{e}_{b}(\epsilon),
\end{equation}
it is easily shown that Eq.\ (2) is expressible as
\begin{equation}
\partial e^{b}(\epsilon)/\partial \epsilon = - e^{c}(\epsilon)
[\nabla_{c}a^{b}]
\end{equation}
where $\nabla_{c}a^{b}$ are the orthonormal components of the covariant
derivative of $\bold{a}(\epsilon)$ on the spacetime $(M,g(\epsilon))$.
\end{description}

That completes our definition.  To guarantee existence, one may assume
that $M$ is compact (which includes the possibility that it is a compact
subspace of another given manifold).  In any case, we shall grant that the
solution of Eqs.\ (2) and (3) over a non-trivial interval $J$ exists and
is $\bold{C}^{\infty}$.  We point out that the above definition and all
conclusions of this paper are also applicable if the prescribed
orthonormal components $a^{b}$ are suitably chosen functions of $\epsilon$
as well as of the spacetime points.  We shall (following \gurses) include
this generalization in the type (b) category.

\section{Neighborhood-Isometric Spacetimes}
\begin{description}
\item[Definition:]  For any set $X$, let $1_{X}$ denote the function whose
domain is $X$ such that $1_{X}(x) = x$ for all $x$ in $X$.  In other
words $1_{X}$ is the identity map on $X$.  For any function $f$, note that
$f \circ 1_{X}$ is the restriction of $f$ to $X$.
\item[Definition:]  Let $(M,g)$ and $(M,h)$ be any spacetimes and
$\bold{x}$ be any point in $M$.  Suppose there exist neighborhoods $V,W$
of $\bold{x}$ and a diffeomorphism $\phi$ of $V$ onto $W$ such that
$\phi^{*}(h \circ 1_{W}) = g \circ 1_{V}$.  Then we shall say that $(M,g)$
and $(M,h)$ are {\em neighborhood-isometric at} $\bold{x}$ and write
$$(M,g)<\mbox{ni},\bold{x}>(M,h)$$
\end{description}
The proof of the following theorem is a pleasant exercise.
\begin{description}
\item[Theorem:]  For given $\bold{x} \in M$, $<\mbox{ni},\bold{x}>$ is an
equivalence relation.
\item[Definition:]  We shall say that $(M,g)$ and $(M,h)$ are
{\em neighborhood-isometric} and write
$$(M,g)<\mbox{ni},M>(M,h)$$
if $(M,g)<\mbox{ni},\bold{x}>(M,h)$ for all $\bold{x}$ in $M$.
\end{description}

It is clear that $<\mbox{ni},M>$ is an equivalence relation.  The key
theorem of this paper can now be simply expressed as follows in terms of
the notations employed in our definition of a type (b) transformation.
\begin{description}
\item[Theorem:]  For all $\epsilon$ in $J$,
\begin{equation}
(M,g(\epsilon)) <\mbox{ni},M> (M,g(0)).
\end{equation}
\end{description}
The proof of the above theorem will be simple once we have proved a lemma
which will be formulated in the next section.  This lemma involves an
explicit construction of the diffeomorphism.

\section{Formulation of a Lemma}
\begin{description}
\item[Definitions:]  For any given point $(\epsilon_{0},\bold{x}_{0})$ in
$J \times M$, let $j_{0} \subset J$ be any connected open neighborhood of
$\epsilon_{0}$ (in the topology of $J$ relative to $R^{1}$) and let
$U \subset M$ be any neighborhood of $\bold{x}_{0}$ such that there exists
a chart $\sigma$ whose domain is $U$.  Also, let $X$ be the range of
$\sigma$ and $x := (x^{1},x^{2},x^{3},x^{4}) = \sigma(\bold{x})$ for all
$\bold{x}$ in $U$.
\item[Definitions:]  Restricting $j_{0}$ and $U$ if necessary to guarantee
existence, and letting $\epsilon$ be any point in $j_{0}$, we define
$f(\epsilon)$ to be that one-one function whose domain is $X$, whose
range is
\begin{equation}
Y(\epsilon) := [\mbox{range of $f(\epsilon)$}] \subset R^{4}
\end{equation}
and which satisfies the familiar flow equation $(\beta=1,2,3,4)$
\begin{equation}
\frac{\partial f(\epsilon,x)}{\partial \epsilon}
+ a^{\beta}(\epsilon,x) \frac{\partial f(\epsilon,x)}{\partial x^{\beta}}
= 0
\end{equation}
and the initial condition
\begin{equation}
f(\epsilon_{0}) = 1_{X} := \mbox{identity map on $X$} \; ,
\end{equation}
where $f(\epsilon,x) := f(\epsilon)(x)$ and where $a^{\beta}(\epsilon,x)$
are the components of $\bold{a}(\epsilon,\bold{x}) :=
\bold{a}(\epsilon)(\bold{x})$ relative to the chart $\sigma$.  Let $y$ be
any point in $Y(\epsilon)$ and
\begin{equation}
h(\epsilon) := [f(\epsilon)]^{-1} \; , \; h(\epsilon,y) :=
h(\epsilon)(y) \; ,
\end{equation}
whereupon it is easily shown that Eq.\ (9) is equivalent to the equation
$(\mu=1,2,3,4)$
\begin{equation}
\partial h^{\mu}(\epsilon,y)/\partial \epsilon =
a^{\mu}(\epsilon,h(\epsilon,y)).
\end{equation}
\end{description}
{}From Eqs.\ (8) and (10), $Y(\epsilon_{0}) = X$.  So we can and we do
further restrict $j_{0}$, if necessary, so that there exists an open
set ${\cal Y} \subset R^{4}$ which satisfies
\begin{equation}
\sigma(\bold{x}_{0}) \in {\cal Y} \subset Y(\epsilon)
\mbox{ for all $\epsilon$ in $j_{0}$} .
\end{equation}
Then the sets defined below are not empty.
\begin{description}
\item[Definitions:]  For all $\epsilon$ in $j_{0}$, let
\begin{equation}
\tau := (1_{{\cal Y}}) \circ \sigma \; , \;
\mu(\epsilon) := 1_{{\cal Y}} \circ f(\epsilon) \circ \sigma \; ,
\end{equation}
which are clearly charts that both have the range ${\cal Y}$ and
that have the domains
\begin{equation}
V := \{ \bold{x} \in U : \sigma(\bold{x}) \in {\cal Y} \}
\end{equation}
and
\begin{equation}
W(\epsilon) := \{ \bold{x} \in U : f(\epsilon,\sigma(\bold{x}))
\in {\cal Y} \}
\end{equation}
respectively.  For all $\epsilon$ in $j_{0}$, let
\begin{equation}
\phi(\epsilon) := [\mu(\epsilon)]^{-1} \circ \tau
\end{equation}
which maps $V$ onto $W(\epsilon)$.
\item[Lemma:]  The diffeomorphism $\phi(\epsilon)$ satisfies
\begin{equation}
[\phi(\epsilon)]^{*} [g(\epsilon) \circ 1_{W(\epsilon)}] =
g(\epsilon_{0}) \circ 1_{V} \mbox{ if $\epsilon$ is in
$j_{0}$.}
\end{equation}
\end{description}
Note that $V$ and $W(\epsilon)$ are both neighborhoods of
$\bold{x}_{0}$.  So the above lemma implies that, for all $\epsilon$ in
$j_{0}$, $(M,g(\epsilon))$ and $(M,g(\epsilon_{0}))$ are
neighborhood-isometric at $\bold{x}_{0}$.

\section{Proof of the Lemma}
An alternative form of Eq.\ (6) is as follows:
\begin{equation}
\frac{\partial e^{b}(\epsilon)}{\partial \epsilon} = - {\cal
L}_{\bold{a}(\epsilon)} e^{b}(\epsilon) + [\bold{a}(\epsilon)
\Gamma_{cd}(\epsilon)] e^{d}(\epsilon) \eta^{cb} \; ,
\end{equation}
where $\bold{a}(\epsilon) \Gamma_{cd}(\epsilon)$ is the value of the
linear functional $\Gamma_{cd}(\epsilon)$ corresponding to the vector
$\bold{a}(\epsilon)$.  The second term on the right side of Eq.\ (19)
represents (when multiplied by $\delta \epsilon$) an infinitesimal
Lorentz transformation of $\{ e^{b}(\epsilon) \}$ and cannot, therefore,
contribute to the variation of the metric tensor with respect to
$\epsilon$.  In fact, a brief calculation employing Eq.\ (4) yields
\begin{equation}
\frac{\partial g(\epsilon)}{\partial \epsilon} =
- {\cal L}_{\bold{a}(\epsilon)} g(\epsilon) \; .
\end{equation}
The theorem of this paper, and even a much stronger form of the theorem,
would now be self-evident if $\bold{a}(\epsilon)$ were independent of
$\epsilon$.  However, the fact that $\bold{a}(\epsilon)$ does generally
depend on $\epsilon$ and is determined in an intricate way by the
prescribed functions at $\epsilon=0$ has led to some doubts as indicated
by our opening remarks.  Therefore, the pedestrian proof which we are
using in this paper may be appreciated by at least some readers.  We
continue by defining a family of real-valued functions
$g_{\mu\nu}(\epsilon)$ and $\gamma_{\alpha\beta}(\epsilon)$.
\begin{description}
\item[Definitions:]  Let $g_{\mu\nu}(\epsilon)$ have the domain $X$ and be
the components of $g(\epsilon)$ relative to the chart $\sigma$.  Let
$\gamma_{\alpha\beta}(\epsilon)$ have the domain $Y(\epsilon)$ and be the
components of $g(\epsilon)$ relative to the chart $f(\epsilon) \circ
\sigma$.  Let
$$g_{\mu\nu}(\epsilon,x) := g_{\mu\nu}(\epsilon)(x) \; , \;
\gamma_{\alpha\beta}(\epsilon,y) := \gamma_{\alpha\beta}(\epsilon)(y) \; .
$$
\end{description}

Then,at each $y$ in $Y(\epsilon)$,
\begin{equation}
\gamma_{\alpha\beta}(\epsilon,y) = g_{\mu\nu}(\epsilon,h(\epsilon,y))
\frac{\partial h^{\mu}(\epsilon,y)}{\partial y^{\alpha}}
\frac{\partial h^{\nu}(\epsilon,y)}{\partial y^{\beta}} \; .
\end{equation}
Furthermore, Eq.\ (20) becomes relative to $\sigma$:
\begin{equation}
\frac{\partial g_{\mu\nu}(\epsilon,x)}{\partial \epsilon}
+ a^{\alpha}(\epsilon,x) \frac{\partial g_{\mu\nu}(\epsilon,x)}
{\partial x^{\alpha}}
+ \frac{\partial a^{\alpha}(\epsilon,x)}{\partial x^{\mu}}
g_{\alpha\nu}(\epsilon,x)
+ \frac{\partial a^{\alpha}(\epsilon,x)}{\partial x^{\nu}}
g_{\alpha\mu}(\epsilon,x) = 0 \; .
\end{equation}
Equations (11), (12), (21) and (22), and the chain rule, now imply:
\begin{equation}
\partial \gamma_{\alpha\beta}(\epsilon,y)/\partial \epsilon = 0 .
\end{equation}
So, upon setting $\epsilon=\epsilon_{0}$ in the left side of Eq.\ (21) and
taking into account the facts that the domains of
$\gamma_{\alpha\beta}(\epsilon)$ and $\gamma_{\alpha\beta}(\epsilon_{0})
=g_{\alpha\beta}(\epsilon_{0})$ are $Y(\epsilon)$ and $Y(\epsilon_{0})=
X$, respectively, we obtain from Eqs.\ (13) and (23):
\begin{equation}
g_{\alpha\beta}(\epsilon_{0},y) = g_{\mu\nu}(\epsilon,h(\epsilon,y))
\frac{\partial h^{\mu}(\epsilon,y)}{\partial y^{\alpha}}
\frac{\partial h^{\nu}(\epsilon,y)}{\partial y^{\beta}}
\mbox{ for all $\epsilon$ in $j_{0}$ and $y$ in ${\cal Y}$.}
\end{equation}
Employing the definitions (14), (15) and (16), we see that Eq.\ (24) is
the component form of the pullback equality
\begin{equation}
(\tau^{-1})^{*}[g(\epsilon_{0}) \circ 1_{V}]
= [\mu(\epsilon)^{-1}]^{*}[g(\epsilon) \circ 1_{W(\epsilon)}] \; .
\end{equation}
Equation (18) then follows from Eqs.\ (17) and (25).  That completes the
proof of the lemma.

\section{Proof of the Theorem}
We merely sketch the proof since the reader can fill in the details
without difficulty.  Consider any given number $\epsilon$ in $J$ and let
$|0,\epsilon|$ be the closed interval with endpoints $0,\epsilon$.  Let
$\bold{x}$ be any point in $M$.  From the lemma, every number
$\epsilon_{0}$ in $J$ can be covered by at least one open interval $j_{0}$
such that
$$(M,g(\epsilon')) <\mbox{ni},\bold{x}> (M,g(\epsilon_{0}))$$
for all $\epsilon'$ in $j_{0}$.  The Heine-Borel covering theorem then
implies that there exists a finite sequence of numbers $\epsilon_{1},
\dots, \epsilon_{A}, \ldots, \epsilon_{N}$ and open subintervals
$j_{1}, \ldots, j_{A}, \ldots, j_{N}$ of $J$ such that $\epsilon_{A}$
lies on $|0,\epsilon|$, $j_{A}$ covers $\epsilon_{A}$, $j_{A}$ and
$j_{A+1}$ overlap for $A=1,\ldots,N-1$, the union of the intervals $j_{A}$
covers $|0,\epsilon|$ and
$$(M,g(\epsilon')) <\mbox{ni},\bold{x}> (M,g(\epsilon_{A}))$$
for all $\epsilon'$ in $j_{A}$.  Therefore, since $<\mbox{ni},\bold{x}>$
is an equivalence relation, we infer that $(M,g(\epsilon))$ and $(M,g(0))$
are neighborhood-isometric at $\bold{x}$.  However, $\epsilon$ and
$\bold{x}$ were arbitrarily chosen members of $J$ and $M$, respectively,
so the theorem [Eq.\ (7)] follows.

\section{Discussion}
In retrospect, the premise that $(M,g(\epsilon))$ is a vacuum spacetime is
never used in the proof and is superfluous for this paper.  Premises
respecting the matter tensor are irrelevant both for the definition of the
type (b) transformation and for the validity of the theorem.  This fact
strongly supports but does not prove the opinion of Torre and Anderson
that their generalized diffeomorphisms are physically trivial\cite{2} or,
as we prefer to express it, that spacetimes which are related by a type
(b) transformation are physically indistinguishable.  We shall now review
this opinion.

It is certain, from the theorem proved in this paper, that every point of
the manifold $M$ can be covered by 1-parameter families of neighborhoods
$V(\epsilon)$ and $W(\epsilon)$ such that the spacetimes
$(V(\epsilon),g(0)|_{V(\epsilon)})$ and
$(W(\epsilon),g(\epsilon)|_{W(\epsilon)})$ are physically equivalent for
all $\epsilon$.  This dashes any hope that type (b) transformations can be
used to generate new exact analytic solutions of the Einstein field
equations {\em regardless of the premises made concerning the matter
tensor}.  However, we hesitate to go beyond the foregoing statements.

We propose that any given spacetimes $(M,g)$ and $(N,h)$ are {\em
physically indistinguishable} if and only if they are isometric\cite{5}
(and not just neighborhood-isometric) or have extensions which are
isometric.  (These extensions need not be maximal.)  Therefore, to prove
that $(M,g(\epsilon))$ and $(M,g(0))$ are physically indistinguishable
{\em on more than just a local level}, one must prove that they are
isometric or have isometric extensions.  Initial efforts in that direction
indicate that the task of proving this or of finding a counter-example may
not be trivial.  Nor do we have any reason to believe that the conjectured
theorem is true.  There are three related research paths which may be
helpful in resolving the issue:
\begin{enumerate}
\item One can focus attention on analytic spacetimes and analytic type (b)
transformations.
\item One can test the validity of the conjectured theorem for
2-dimensional Lorentzian manifolds.
\item One can investigate the effects on type (b) transformations of
critical points of $\bold{a}(0)$.
\end{enumerate}

We leave the above ventures for interested readers since we intend to move
on to other kinds of symmetries.

\section*{Acknowledgement}
This work was supported in part by grant PHY-92-08241 from the National
Science Foundation.

\section*{References:}
\begin{enumerate}
\item
M.\ \gurses, Phys.\ Rev.\ Lett.\ {\bf 70}, 367 (1993).
\item
C.\ Torre and I.\ Anderson, Phys.\ Rev.\ Lett. {\bf xx}, xxx (1993).
\item
This can be seen by inserting Eq.\ (16) of our paper into Eq.\ (14)
and then comparing the result with Eq.\ (6) of Ref.\ 2.
\item
C.\ G.\ Torre (private communication).
\item
This concept is based on the concept of isometric Riemannian or
semi-Riemannian manifolds as defined by R.\ K.\ Sachs and H.\ Wu,
{\em General Relativity for Mathematicians} (Springer-Verlag,
New York, 1977) p.\ 4.  The concept of neighborhood-isometry is
our own invention, to the best of our knowledge.  It is
not to be confused with the concept of local isometry as defined
by Sachs and Wu on p.\ 4 of their book.
\item
F.\ J.\ Ernst and I.\ Hauser, submitted to J.\ Math.\ Phys.
\end{enumerate}

\end{document}